\documentclass[runningheads,a4paper]{llncs}
\usepackage{booktabs} 
\usepackage{multirow}
\usepackage{subfig}
\usepackage{algorithm}  
\usepackage{algorithmic}
\usepackage{mathtools}
\usepackage{caption}
\usepackage{t1enc}

\usepackage{amssymb}
\usepackage{graphicx}
\usepackage{epstopdf}

\usepackage{url}
\urldef{\mailsa}\path|{gaoyan}@software.ict.ac.cn,|
\urldef{\mailsb}\path|{guojiafeng, lanyanyan,lhm}@ict.ac.cn|

\begin{document}
\title{Dynamic-K Recommendation with Personalized Decision Boundary}

\author{Yan Gao \and Jiafeng Guo\and Yanyan Lan\and Huaming Liao}
\institute{CAS Key Lab of Network Data Science and Technology
Institute of Computing Technology, Chinese Academy of Sciences\\
\mailsa\\
\mailsb}

\maketitle
\begin{abstract}
In this paper, we investigate the recommendation task in the most common scenario with implicit feedback (e.g., clicks, purchases). State-of-the-art methods in this direction usually cast the problem as to learn a personalized ranking on a set of items (e.g., webpages, products). The top-N results are then provided to users as recommendations, where the N is usually a fixed number pre-defined by the system according to some heuristic criteria (e.g., page size, screen size). There is one major assumption underlying this fixed-number recommendation scheme, i.e., there are always sufficient relevant items to users' preferences. Unfortunately, this assumption may not always hold in real-world scenarios. In some applications,  there might be very limited candidate items to recommend, and some users may have very high relevance requirement in recommendation. In this way, even the top-1 ranked item may not be relevant to a user's preference. Therefore, we argue that it is critical to provide a dynamic-K recommendation, where the K should be different with respect to the candidate item set and the target user. We formulate this dynamic-K recommendation task as a joint learning problem with both ranking and classification objectives. The ranking objective is the same as existing methods, i.e., to create a ranking list of items according to users' interests. The classification objective is unique in this work, which aims to learn a personalized decision boundary to differentiate the relevant items from irrelevant items. Based on these ideas, we extend two state-of-the-art ranking-based recommendation methods, i.e., BPRMF and HRM, to the corresponding dynamic-K versions, namely DK-BPRMF and DK-HRM. Our experimental results on two datasets show that the dynamic-K models are more effective than the original fixed-N recommendation methods. 

\end{abstract}

\section{Introduction}

Recommender systems have been widely used in many applications, such as Amazon, YouTube and so on. In this paper, we address the most common recommendation scenario with implicit feedbacks, e.g., clicks or purchases from users. Most methods in this direction cast the problem as to learn a personalized ranking on a set of items\cite{karatzoglou2013learning}, e.g., webpages or products. The top-N ranked items are then provided to users as recommendations\cite{adomavicius2005toward,oard1998implicit,ricci2011introduction}, where N is usually a fixed number pre-defined by the system according to the recommendation space (e.g., page size, screen size) or some heuristic criteria.  

There is an underlying assumption for such fixed-N recommendation scheme, i.e., there are always sufficient relevant items to users' preferences. Unfortunately, this assumption may not always hold in real-world scenarios. For example, if one aims to recommend newly uploaded papers on arxiv\footnote{https://arxiv.org/} for academic readers everyday, he/she may face the following two problems. Firstly, there might be very limited new papers updated on arxiv everyday. Secondly, some academic readers might have very high relevance requirement in recommendation since paper reading is time-consuming. In this way, even the top-1 ranked paper may not be appropriate to be recommended to such readers on some day. However, by using a fixed-N recommendation scheme, each reader will constantly receive N recommended papers, and it is very likely to annoy the readers in this situation.

To avoid the above problem, we argue that it is critical to provide a dynamic-K recommendation, where the K should be different with respect to the candidate item set and the target user. If there are sufficient relevant items or if the user likes receiving diverse recommendations (i.e., relatively low relevance requirement), the K could be large. In contrary, if there are limited relevant items or if the user likes receiving recommendations unless they are highly relevant, the K should be small or even no recommendation should be provided sometimes. Ideally, a good recommender system should be able to learn from users' implicit feedbacks to present such dynamic-K recommendations for different candidate item set and target users. 

In this work, we formulate this dynamic-K  recommendation task as a joint learning problem with both ranking and classification objectives. Specifically, the ranking objective is the same as many existing ranking-based methods, i.e., to create a ranking list of items according to users' interests. The classification objective, which is unique in this work, aims to learn a personalized rather than a global decision boundary to differentiate the relevant items from irrelevant items. We apply the above joint learning idea over two state-of-the-art ranking-based recommendation methods, i.e., BPRMF and HRM, and extend them to the corresponding dynamic-K versions, namely DK-BPRMF and DK-HRM, respectively.

We conduct empirical experiments over two publicly available datasets, i.e., a transaction dataset named Ta-Feng and a movie recommendation dataset named MovieLens. 
Our experimental results show that the dynamic-K models are more effective than the corresponding fixed-N recommendation methods as well as those existing hybrid recommendation methods.

\section{Related Work}
Many methods have been developed in literature to build implicit feedback recommendation systems. With repsect to the objective function, the recommendation methods can be further divided into three folds, namely ranking-based methods, classification-based methods and hybrid methods.

\textbf{Ranking-based recommendation methods}, which aim to correctly rank items rather than to correctly predict their ratings have demonstrated good performance for top-N recommendation system.
Ranking-based recommendation methods can be categoried into two folds, i.e., pair-wise approach and list-wise approach. 
Pair-wise approach aims to optimize the pair-wise loss. For example, Rendle et al.\cite{rendle2009bpr} and Aiolli et al.\cite{aiolli2014convex} optimized AUC score; Yun et al.\cite{yun2014ranking} explored the connection between the Discounted Cumulative Gain(DCG) and the binary classification to change the ranking problem into binary classification problems; Park et al.\cite{park2015preference} proposed a large-scale collaborative ranking method to minimize the ranking risk in the reconstructed recommendation matrix. List-wise approach optimizes the preference of each user to a list of items. An important branch in this category is designed to directly optimize evaluation merics, such as Mean Average Precision(MAP), Mean Reciprocal Rank(MRR) and Normalized Discounted Cumulative Gain(NDCG), which are usually list-wise ranking metrics. Typical methods include TFMAP\cite{shi2012tfmap}, CLiMF\cite{shi2012climf} and CofiRank\cite{weimer2007maximum}.
Different types of ranking approaches have different strengths in producing the ranking list, but all these ranking-based methods are proposed under the fixed-N recommendation scenario.

\textbf{Classification-based recommendation methods} attempt to predict whether a user would like to interact with an item. Mnih et al.\cite{mnih2011learning} provide a probabilistic framework for the implicit case where they model the probability of a user choosing an item according to a normalized exponential function. They avoid linear time computation and approximate the normalization to the distribution by traversing a tree structure. Goplan et al.\cite{gopalan2013scalable} introduced a factorization model that factorizes users and items by Poisson distribution. More recently, Johnson \cite{johnson2014logistic} proposed a new probabilistic framework for the implicit case in which they model the probability of a user choosing an item by a logistic function. Nevertheless, these methods do not learn a personalzied classifier for each user, thus can't be used for dynamic-K recommendation scenario. 


\textbf{Hybrid recommendation methods} try to optimize several objectives simultaneously which can integrate the complementary strengths of different types of losses. Zhao et al.\cite{zhaoimproving} propose a personalized top-N recommendation approach that minimizes a combined heterogeneous loss based on linear self-recovery models. The heterogeneous loss integrates the strengths of both pair-wise ranking loss and point-wise recovery loss to provide more informative recommendation predictions. 
Compared with their heterogeneous loss based on linear self-recovery models, we formulate our dynamic-K recommendation task as a joint learning problem with both ranking and classification objectives. Specifically, our classification objective aims to learn a personalized decision boundary for each user to differentiate the relevant items from irrelevant items.


\section{Our Approach}
In this section, we first introduce the problem formalization of recommendation with implicit feedback. We then describe the joint learning approach in detail. After that, we show how it can be applied to extend two state-of-the-art ranking-based recommendation methods, i.e., BPRMF and HRM, to the corresponding dynamic-K versinons, namely DK-BPRMF and DK-HRM.

\subsection{Problem Definition}

Let $U$ be a set of users and $I$ be a set of items. In our scenario, implicit feedback $X\subseteq U\times V$ is available. Each instance $(u,i)\in X$ is a pair which means an interaction between user $u$ and item $i$. Formally, we define the set of items with which user $u$ has interactions is ${ B }_{ u }^{ + }:=\{ i\in I|(u,i)\in X\}$. With the learned model, we generate a list of all candidate items for each user. The top K ranked items are then provided to users as recommendations, where K is different with repsect to the candidate item set and the target user. That is called dynamic-K recommendation system. 

\subsection{Joint Learning Approach}

In this work, we formulate the dynamic-K recommendation task as a joint learning problem with both ranking and classification objectives. Specifically, the ranking objective could use any existing ranking-based methods, i.e., to create a ranking list of items according to user's interests. Since we are dealing with implicit feedback data, we follow the rationale from Rendle et al.\cite{rendle2009bpr} assuming that user $u$ prefers item $i$ over item $j$ if $i\in { B }_{ u }^{ + }\wedge j\notin { B }_{ u }^{ + }$. 
Formally, we define the set $P$ as the set of tuples $\{(u,i,j)\}$ selected from dataset $X$ as follows: $P=\{ (u,i,j)|i\in { B }_{ u }^{ + }\wedge j\notin { B }_{ u }^{ + }\} $. Therefore, the ranking objective can be formazlied as:
\begin{equation}
{ L(P;\theta) }_{ rank }=\sum _{ (u,i,j)\epsilon P }{ \ell({ s }_{ u,i }\succ { s }_{ u,j };\theta ) } 
\end{equation}
where $\theta$ represents the parameter vector of an arbitrary model(e.g. matrix factorization). $s_{u,i}$, $s_{u,j}$ means the predicted score of item $i$ to user $u$ and item $j$ to user $u$ respectively. $\ell(\cdot)$ can be any arbitrary ranking loss function.

The classification objective, which is unique in this work, aims to learn a personalized decision boundary to differentiate the relevant items from irrelevant items. Intuitively, if the predicted score of the item is above user's decision boundary, the item is relevant. If the predicted score is under user's decision boundary, the item is irrelevant. We use a variable called $margin_{ui}$ to measure the confidence of the predicted class of item $i$ to user $u$(i.e., if item $i$ is relevant or irrevelant to user $u$):
\begin{equation}
margin_{ui}=y_{ui}(s_{u,i}-t_u)
\end{equation}
where $y_{ui}\in\{-1, 1\}$ is the target class, $t_u$ is the personazlied decision boundary for each user. $s_{u,i}$ is the predicted score of item $i$ to user $u$ with an arbitary model(e.g. matrix factorification). $Margin_{ui} <0$ indicates item $i$ is misclassified to user $u$, while $margin_{ui} >0$ indicates item $i$ is correctly classified to user $u$. $Margin_{ui}$ represents the "margin of safety" by which the prediction for item $i$ to user $u$ is correct. Furthermore, we assume the personalized decision boundary is regularized by a global constraint to avoid over-fitting. 
For implicit feedback data, 
we define the dataset $D$ as a set of tuples $\{(u,i,y_{ui})\}$ selected from $U\times I$, and $y_{ui}$ is defined as follows: 
\begin{equation}
{ y }_{ ui }=\begin{cases} 1\quad (u,i)\in X\\ -1\quad (u,i)\notin X\end{cases}
\end{equation}
Therefore, the learning objective in classification is to find the best $\theta$ and $t_u$ to minimize classification loss over the training data $D$ as following:
\begin{equation}
{ L(D;\theta, t_u) }_{ cf }=\sum _{ (u,i,y_{ui})\in D }^{  }{ \ell({ margin }_{ ui };\theta) } +{ \lambda  }_{ t }R({ t }_{ u }-t)
\end{equation}
where $\ell(\cdot)$ is defined as a loss function of the margin for each data,
$t$ is the global constrain we set for $t_u$ and $R( \cdot )$ is a regularizer(typically L2 or L1). $\lambda_t>0$ is a co-efficient controling the regularization strength. 
As noted above, any classification loss function could be applied in this framework. In particular, we explored the use of logisitc loss and hinge loss in preliminary experiments, but found that logistic loss performed better on the data sets in this paper. Logistic loss is as following:
\begin{equation}
\quad { \ell }_{ log }=ln(1+{ e }^{ -margin_{ui} })
\end{equation}


Finally, by combining (1) and (4), we obtain our joint learning approach as follows: 
\begin{equation}
{ L }_{ hybrid }=\alpha { L }_{ cf }+(1-\alpha ){ L }_{ rk }
\end{equation}
where the parameter $\alpha \in \left[ 1,0 \right]$ denotes the trade-off between optimizing ranking loss and classification loss. Note that by setting $\alpha$ = 1, the model reduces to a classification-based model; By setting $\alpha$ = 0, we obtain a ranking-based model.

A direct algorithm for optimizing the joint objective function would enumerate the full set $P$ of candidate pairs. Because $\left|P\right|$ is quadratic in $\left|D\right|$, this would be intractable for large-scale data sets. Instead, following similar idea with Sculley\cite{sculley2010combined}, we take the sampling approach from $P$ rather than constructing $P$ explicitly.
Algorithm 1 gives a method for efficiently solving the joint learning optimization problem using stochastic gradient descent(SGD)\cite{shalev2007pegasos}.



\begin{algorithm}[h]  

\caption{Joint Learning Approach.
Given: tradeoff parameter $\alpha$, regularization parameter $\lambda_t$, iterations $N$}  
\begin{algorithmic}[1]  
\STATE $\theta,t_u\leftarrow \oslash,t$

\FOR{$n=1$ to $N$}  
\STATE pick z uniformly at random from $\left[0, 1\right]$ 
\IF{$z < \alpha$}   
\STATE $(u, i, y_{ui}) \leftarrow RandomExample(D)$ 
\STATE $L_{cf}=\ell(margin_{ui}) +{ \lambda  }_{ t }R({ t }_{ u }-t)$
\STATE $\theta ^{ n }\leftarrow StochasticGradientStep(\theta^{ n-1 },\frac { \partial { L }_{ cf } }{ \partial \theta  } ,\eta )$
\STATE $t_u^{ n }\leftarrow StochasticGradientStep(t_u^{ n-1 },\frac { \partial { L }_{ cf } }{ \partial t_u  } ,\lambda_t, \eta )$
\ELSE   
\STATE $((u,i,j))\leftarrow RandomCandidatePair(P)$
\STATE $L_{rk}=\ell({ s }_{ u,i }\succ { s }_{ u,j };\theta )$
\STATE $\theta^n \leftarrow StochasticGradientStep(\theta^{n-1},\frac { \partial { L }_{ rk } }{ \partial \theta  } ,\eta)$
\ENDIF   
\ENDFOR 

\label{code:recentEnd}  
\end{algorithmic}  
\end{algorithm}  

In prediction stage, with the learned model, the dynamic-K recommendation with our joint learning approach is as following. Given a user $u$ and an itemset $I$, for each candidate item $i\in I$, we calculate the predicted score $s_{ui}$. We then rank the items according to their predicted score, and select the top K results 
above the threshold $t_u$ as the final recommendation results.

\subsection{Implementation of the Joint Approach}
We can see that the proposed joint approach is a general framework which can be appied to any existing learning to rank methods. In the following, we adopt two state-of-the-art ranking-based recommendation methods, i.e., BPRMF and HRM, and extend them to the corresponding dynamic-K versions, namely DK-BPRMF and DK-HRM, respectively.
\subsubsection{DK-BPRMF}
Bayesian personalized ranking maxtrix factoriaztion(BPRMF) is a matrix factorization model with BPR\cite{rendle2009bpr} as optimize criteria. With matrix factorization, the target matrix $X$ is approximated by the matrix product of two low-rank matrices $P:|U|\times f$ and $Q:|I|\times f$:
\begin{equation}
\hat{X}:=PQ^t
\end{equation}
where $f$ is the dimensionality/rank of the approximation. Thus the prediction formula can also be written as:
\begin{equation}
{ \hat { x }  }_{ ui }=<{ p }_{ u },{ q }_{ i }>=\sum _{ f=1 }^{ F }{ { p }_{ uf }\cdot { q }_{ if } } 
\end{equation}

BPR is a state-of-the-art pairwise ranking framework for the implicit feedback data, specifically, it use the maximum posterior estimator that is derived from a Bayesian analysis of the problem. Therefore, the ranking objective can be written as:
\begin{equation}
{\ell}_{rk}=\sum_{(u,i,j)\in P} \sigma(x_{ui}-x_{uj})+\lambda _\theta\left \| \theta \right \|^2 
\end{equation}
where $\sigma$ is the logistic sigmoid $\sigma (x):=\frac { 1 }{ 1+{ e }^{ -x } }$, and $\lambda_{\theta}$ is a regularization parameter that controls the complexity of the model. According to our joint approach, the classification objective with logistic loss can be written as :
\begin{equation}
{ \ell}_{ cf }=\sum _{ (u,i,y_{ ui })\in D }ln(1+e^{-y_{ui}(x_{ui}-t_u)}) +\lambda _t \left \| t_u - t \right \|^2  \\
\end{equation}
Therefore, the dynamic-K version of BPRMF is:
\begin{equation}
\ell_{DK\text{-}BPRMF} = \alpha \cdot \ell_{cf} + (1-\alpha)\cdot \ell_{rk} 
\end{equation}

\subsubsection{DK-HRM}

HRM is a state-of-the-art model for next basket recommendation which can capture both sequential behavior and users' general taste, and meanwhile model complicated interactions among these factors in prediction.

For each user $u$, a purchase history $T^u$ is given by ${ T }^{ u }:=({ T }_{ 1 }^{ u },{ T }_{ 2 }^{ u },...,{ T }_{ { t }_{ u }-1 }^{ u })$, where ${ T }_{ t }^{ u }\subseteq I$, $t\in \left[ 1,{ t }_{ u }-1 \right] $. 
Given this history, the task of next basket recommendation is to recommend items that user $u$ would probably buy at the next visit.
We define the set $H$ as the set of quaternion $\{(u,t,i,j)\}$ selected from dataset as follows: $H=\{ (u,t,i,j)|i\in { T }_{ t }^{ u }\wedge j\notin { T }_{ t }^{ u }\} $. The set $D$ is a set of quaternion $\{(u,t,i,y_{u,t,i})\}$ selected from $U\times T\times I$, and $y_{u,t,i}$ is defined as follows: 
\begin{equation}
{ y }_{ u,t,i }=\begin{cases} 1\quad i\in { T }_{ t }^{ u } \\ -1\quad i\notin { T }_{ t }^{ u } \end{cases}
\end{equation}

Given a user $u$ and his/her two consecutive transactions $T_{t-1}^{u}$ and $T_{t}^{u}$, HRM defines the score of buying next item $i$ as following:
\begin{equation}
 { x }_{ u,t,i }={ v }_{ i }\cdot { v }_{ u,t-1 }^{ Hybrid } 
\end{equation}
where $v_i$ denotes the representation of item $i$ and ${ v }_{ u,t-1 }^{ Hybrid }$ denotes the hybrid representation obtained from the hierarchical aggregation according to $T^{u}_{t-1}$.

In the learning process, the ranking objective with logistic loss  is written as follows:
\begin{equation}
\ell_{rk} =\sum_{(u,t,i,j)\in H} \frac { 1 }{ 1+{ e }^{ -(x_{ u,t,i }-x_{ u,t,j })} }
\end{equation}
The classification objective is as follows:
\begin{equation}
{ \ell}_{ cf }=\sum _{ (u,t,i,y_{ u,t,i })\in D }ln(1+e^{-y_{u,t,i}(x_{u,t,i}-t_u)})+\lambda _t \left \| t_u - t \right \|^2  \\
\end{equation}
Finally, the dynamic-K version of HRM is:
\begin{equation}
\ell_{DK\text{-}HRM} = \alpha \cdot \ell_{cf} + (1-\alpha)\cdot \ell_{rk}
\end{equation}

\section{Evaluation}
In this section, we conduct empirical experiments to demonstrate the effectiveness of our proposed joint learning models DK-BPRMF and DK-HRM for dynamic-K recommendation. We first introduce the dataset, baseline methods, and the evaluation metrics employed in our experiments. Then we compare the performance of hyper parameters for dynamic-K recommendation. After that, we compare our DK-BPRMF and DK-HRM to the corresponding fixed-N versions to demonstrate effectiveness.

\subsection{Datasets}
We use two datasets. The Movielens100K dataset and a retail datasets Ta-Feng. 

\begin{itemize}
 \item The Movielens100K dataset, it contains 100,000 ratings (1-5) from 943 users on 1682 movies. Each user has rated at least 20 movies.
 \item The Ta-Feng dataset is a public dataset released by RecSys conference, which covers products from food, office supplies to furniture. It contains 817, 741 trans- actions belonging to 32, 266 users and 23, 812 items. 

\end{itemize}

We first conduct some pre-process on these datasets. For Movielens100K, as we want to solve an implicit feedback task, we removed the rating scores from the dataset. Now the task is to predict if a user is likely to rate a movie. For Ta-Feng dataset, we remove all the items bought by less than 10 users and users that has bought in total less than 10 items. In order to simulate the real world recommendation scenatio, finally, we seprate the training data and testing data by time. For Ta-Feng, the testing set contains only the last transaction of each user, and for Movielens100K, the testing set contains only the last rated movies of each user, while all the remaining interactions are put into the training set. The models are then learned on $S_{train}$ and their predicted personalized ranking is evaluated on the test set $S_{test}$.

\subsection{Evaluation Methodology}
The performance is evaluated for each user $u$ in the testing set. For each recommendation method, we generate a list of K items for each user $u$, denoted by $R(u)$, where $R_{i}(u)$ stands for the item recommended in the $i$-th position. We use the following measures to evaluate the recommendation lists.

\begin{itemize}
 \item F1-score: F1-score is the harmonic mean of precision and recall, which is a widely used measure in recommendation:
\begin{equation}
Precision(R(u), S_{test}^u):={ \frac { \left| R(u)\cap { S }_{ test }^{ u } \right|  }{\left| R(u) \right| }  } 
\end{equation}

\begin{equation}
Recall(R(u), S_{test}^u):={ \frac { \left| R(u)\cap { S }_{ test }^{ u } \right|  }{ \left| { S }_{ test }^{ u } \right|  }  }  
\end{equation}

\begin{equation}
F1\text{-}score:=\frac { 2\times Precision\times Recall }{ Precision+Recall } 
\end{equation}

\item NDCG@k: Normalized Discounted Cumulative Gain (NDCG) is a ranking based measure which takes into account the order of recommended items in the list[11], and is formally given by: 
\begin{equation}
NDCG@k=\frac { 1 }{ { N }_{ k } } \sum _{ j=1 }^{ k }{ \frac { { 2 }^{ I({ R }_{ j }(u)\in { T }_{ { t }_{ u } }^{ u }) }-1 }{ { log }_{ 2 }(j+1) }  } 
\end{equation}
where $I(\cdot·)$ is an indicator function and $N_k$ is a constant which denotes the maximum value of NDCG@k given $R(u)$.

\item Cover-Ratio: In dynamic-K recommendation, Cover-Ratio measures how many users can get recommendation results. 
\begin{equation}
Cover\text{-}Ratio=\frac { \sum { I(\left |R(u)\right |>0) }  }{ |U| } 
\end{equation}
\end{itemize}

\subsection{Baseline and Parameters Setting}
The baselines include
\begin{itemize}
\item LogMF\cite{johnson2014logistic}: LogMF is a probabilistic model for matrix factorization with implicit feedback. 
\item BPRMF\cite{rendle2009bpr}: BPRMF is a matrix factorization model with BPR as optimize criteria. BPR is a state-of-the-art pairwise ranking framework for the implicit feedback data.
\item CRRMF\cite{sculley2010combined}: CRRMF is a matrix factorization model with CRR as optimize criteria. CRR is a hybrid framework combining ranking and regression together.
\item HRM\cite{wang2015learning}: HRM is a state-of-the-art hybrid model on next basket recommendation. Both sequential behavior and users’ general taste are taken into account for prediction and meanwhile modeling complicated interactions among these factors in prediction.

\end{itemize}

For all the models, we fixed the dimension of latent factors in $U$ and $I$ to be 50. In LogMF, similar to Johnson\cite{johnson2014logistic}, for each triple $(u,i,j)\in P$, $i$ represents positive observations, $j$ represents negative observations and we tune the parameter $c$ to balance the positive and negative observations. In CRRMF, for each triple $(u,i,j)\in P$, we regress $i$ and $j$ to 1 and 0 respectively.  In HRM and DK-HRM, we use the average pooling. Parameter tuning was performed using cross validation on the training data for each method. We update parameters utilizing Stochastic Gradient Ddescent(SGD) until converge.  
\begin{figure*}
\setlength{\abovecaptionskip}{0.cm}
\setlength{\belowcaptionskip}{-0.5 cm}
\centering
\includegraphics[height=2.5 in, width=4.4 in]{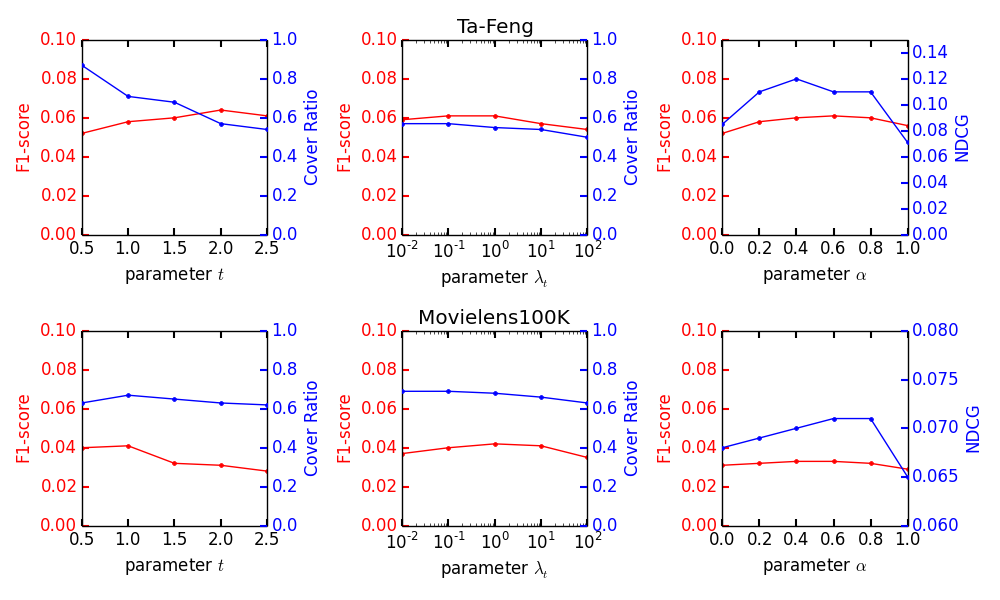}
\caption{Performance comparison among hyper parameters of DK-HRM over two datasets.}
\end{figure*}
\subsection{Comparison among Different Hyper Parameters}

For the proposed method, we have three hyper parameters, $t$, $\lambda_t$ and $\alpha$. Because the limitation of space, we only report the result of DK-HRM over two datasets Ta-Feng and Movielens100K. Similar results can also be obtained from DK-BPRMF. The results over two datasets are shown in Figure 1.

Firstly, we explored how the parameter $t$ affects the performance of DK-HRM. By setting $\lambda_t$ and $\alpha$ to be 1.0 and 0.5 respectively, we selected $t$ from \{0.5, 1.0, 1.5, 2.0, 2.5\}. As we can see, as $t$ increases, F1-score increases but Cover-ratio descreses, indicating that the higher global constraint of personalized decision boundary we set, the higher F1-score we get, but the lower Cover-ratio we get. Take Ta-Feng as an example, when compared with $t=0.5$, the relative improvement of F1-score by $t=2.0$ is around 30\%, and the decrease of Cover-ratio is 35\%. 

Then, we fixed $t$ as 2.0 and compared different value of $\lambda_t$ from \{0.01, 0.1, 1, 10, 100\}. As we can see, with the increase of $\lambda_{t}$, which means the increase of regularization strength, F1-score gradually increases, reaches the peak and then starts to decrease. It indicates that appropriate regularization strength can improve metric performance. Take Ta-Feng as example, F1-score reaches peak when $\lambda_{t}=1.0$, then starts to decrease. 

At last, we fixed $t$ as 2.0 and $\lambda_t$ as 1.0 and selected $\alpha$ from \{0, 0.2, 0.4, 0.6, 0.8, 1.0\}. $\alpha$ denotes the trade-off between optimizing ranking loss and classification loss. As we can see, as $\alpha$ increases, F1 score and NDCG increase first, and then decrease, indicating that the two types of losses can complement each other and achieve the best performance at the same time.


\subsection{Comparison against Baselines}
We further compare our DK-BPRMF and DK-HRM methods to the state-of-the-art baseline methods. Since LogMF, BPRMF, CRRMF and HRM get top-N recommendation results, in order to make a fair comparison between top-N and dynamic-K, we vary the number N and get the F1-score of top-N method from top-1 to top-20, shown in Figure 2. We find that as the number of recommendtion items increases, F1-score increases and then keeps steady. Take Ta-Feng as example, for HRM, F1-score increases to 5.2\% and then keeps steady.
\begin{figure}
\setlength{\abovecaptionskip}{0.cm}
\setlength{\belowcaptionskip}{-0.cm}
\centering
\includegraphics[height=1.8 in, width=4.0 in]{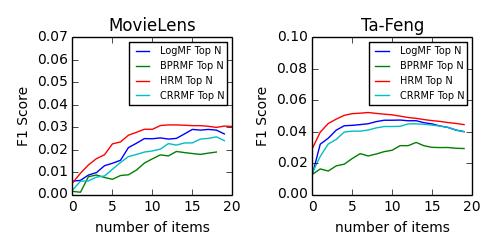}
\caption{The top-N F1-score of the Ta-Feng dataset and MovieLens dataset. The number of recommended items is set 1 to 20 on Ta-Feng and MovieLens.}
\end{figure}

\begin{table*}
\setlength{\abovecaptionskip}{0.2 cm}
\setlength{\belowcaptionskip}{-0 cm}
\caption{Performance comparison of DK-HRM and DK-BPRMF among LogMF, BPRMF, CRRMF and HRM over two datasets. Significant improvement of our model(DK-BPRMF and DK-HRM) with respect to their original baseline methods(BPR and HRM respectively)is indicated as $\textquoteleft$+'(p-value $\le$ 0.05).}
\scriptsize
\centering
  \begin{tabular}{cccccccl}  
  \toprule
\multirow{2}{*}{Models} & \multicolumn{3}{c}{Ta-Feng}   & \multicolumn{3}{c}{MovieLens} &  \\
& F1-score & Cover-ratio& NDCG & F1-score & Cover-ratio & NDCG &  \\
\midrule
BPRMF & 0.033 & 1.0  & 0.075 & 0.018 & 1.0 & 0.067 &\\\hline
LogMF & 0.042 & 1.0  & 0.071 & 0.028 & 1.0  & 0.058 &\\
CRRMF & 0.041 & 1.0 & 0.074 & 0.026 & 1.0  & 0.068 &\\
\textbf{DK-BPRMF} & \textbf{0.048$^+$} & \textbf{0.69} & \textbf{0.11$^+$} & \textbf{0.021$^+$} & \textbf{0.9}  & \textbf{0.135$^+$}\\\hline
HRM & 0.052 & 1.0 & 0.08 & 0.03 & 1.0 & 0.117 &\\
\textbf{DK-HRM} & \textbf{0.06$^+$} & \textbf{0.7}  & \textbf{0.12$^+$} & \textbf{0.035$^+$} & \textbf{0.68} & \textbf{0.12$^+$}\\
\bottomrule
\end{tabular}

\end{table*}

Then, we choose the best performed DK-BPRMF($t$ is set 1.0, $\alpha$ is set 0.5 and $\lambda_t$ is set 1.0) and DK-HRM($t$ is set 2.0, $\alpha$ is set 0.5 and $\lambda_t$ is set 1.0) and the best performed baselines from Figure 2 as the representative for fair comparison. Results over Ta-Feng and Movielens100K are shown in Table 1. Three main obersavations can be drawn from the results. (1) BPRMF performs better in NDCG and LogMF performs better in F1-score respectivly, due to their different optimiaztion goal for ranking and classification respectively. As a hybrid model, CRRMF can achieve both good ranking and classification performance compared with BPRMF and LogMF. (2) HRM outperforms BPRMF, LogMF and CRRMF in all metrics since it can capture both sequential behavior and users' general taste. Take Ta-Feng as example, F1-score and NDCG of HRM reach 5.2\% and 0.08 repectively. (3) Compared with top-N recommenders, DK-BPRMF and DK-HRM for dynamic-K recommendation can consistently outperform their original baseline methods respectively(i.e. BPR and HRM) in terms of F1-score and NDCG over the two datasets. Take the Ta-Feng dataset as an example, when compared with the second best performed baseline method (i.e. HRM), the relative improvement by DK-HRM($t$ set as 2.0,$alpha$ set as 0.5 and $\lambda_{t}$ set as 1.0) is around 33.1\%, 50\%, in terms of F1-score and NDCG, respectively. 

It indicates that our joint learning approach for dynamic-K recommendation is more effective than top-N recommendation methods.

\section{Conclusions}
In this paper, firstly, we argue that it is critical to provide a dynamic-K recommendation where the K should be different with respect to the candidate item set and the target user instead of fixed-N recommendation. Then we formulate this dynamic-K recommendation task as a joint learning problem with both ranking and personazlied classification objectives. We apply the above joint learning idea over two state-of-the-art ranking-based recommendation methods, i.e., BPRMF and HRM, and extend them to the corresponding dynamic-K versions, namely DK-BPRMF and DK-HRM, respectively. Our experimental results show that the dynamic-K models are more effective than the corresponding original fixed-N recommendation methods.

{
\selectfont
\tiny
}

\end{document}